# Parameter Space for Orbits in the Schwarzschild Metric

F.T. Hioe* and David Kuebel

Department of Physics, St. John Fisher College, Rochester, NY 14618

and

Department of Physics & Astronomy, University of Rochester, Rochester, NY 14627, USA

November 23, 2010

**Abstract**

Some features of a parametrized space of orbits in the Schwarzschild geometry are described.

In two recent papers [1,2], we present analytic expressions that describe the trajectories of a non-zero mass particle characterized on a map in a dimensionless parameter space $(e, s)$, where we have called $e$ the energy parameter and $s$ the field parameter, but we have confined our characterization of the trajectories to the region of the space in which $0 \leq e \leq \infty$, and $0 \leq s \leq \infty$. As we pointed out recently [3], this parameter space needs to be extended in order to represent all possible orbits. The enlarged parameter space is $(e^2, s)$ where $-\infty \leq e^2 \leq \infty$, and $0 \leq s \leq \infty$, and we observed that the "extended" space represented by $e^2 < 0$ does not have a Newtonian correspondence. This parameter space appears to have some interesting features which we describe in this note. Some of these features may be related to physical properties of the orbits.

The dimensionless quantities $(e^2, s)$ used to parametrize the orbits are defined by

$$e^2 = 1 + \frac{h^2 c^2 (\kappa^2 - 1)}{(GM)^2} \tag{1}$$

and

$$s = \frac{GM}{hc} \tag{2}$$

where $M$ is the mass of the central object, $\kappa$ is the total energy of the orbiting particle divided by its rest mass energy, and $h$ is the angular momentum per unit rest mass of the particle. The parameter space in our work presented in refs.1 and 2 was confined to $e \geq 0$ and $s \geq 0$. General relativity allows an extended parameter space that includes $e^2 < 0$. As explained in ref.3, let us first note what $e^2 \geq 0$ and $e^2 < 0$ mean physically. From the definitions of $\kappa$ and $h$ and using the 'combined' energy equation [4]

$$\dot{r}^2 + \frac{h^2}{r^2}\left(1 - \frac{\alpha}{r}\right) - \frac{c^2\alpha}{r} = c^2(\kappa^2 - 1) \tag{3}$$

with $\alpha = 2GM/c^2$ as the Schwarzschild radius we find that $e^2 \geq 0$ or $< 0$ respectively imply

$$\dot{r}^2 + \left(r\dot{\phi} - \frac{GM}{r^2\dot{\phi}}\right)^2 \geq or < \frac{2GM}{c^2}r\dot{\phi}^2. \tag{4}$$

In the Newtonian limit the right-hand side of the above inequalities is assumed to be a small quantity (but not zero), and thus while the parameter space $e^2 \geq 0$ has its Newtonian correspondence, the "extended" space characterized by $e^2 < 0$ does not have a Newtonian correspondence.

The parameter $e$ has a well defined Newtonian limit when $e \geq 0$ which is characterized by $s \simeq 0$ as was described in our earlier work. The constant $c^2(\kappa^2 - 1)$ which is $< 0$ for a bound orbit and $\geq 0$ for an unbound orbit, can be identified with $2E_0/m$ in the Newtonian limit, where $E_0$ is the sum of the kinetic and potential energies and is $< 0$ for a bound orbit, and is $\geq 0$ for an unbound orbit, and $m$ is the mass of the particle (which approaches $m_0$). In this limit we therefore obtain

$$e \simeq \left[1 + \frac{2E_0 h^2}{m(GM)^2}\right]^{1/2}, \tag{5}$$

which can be identified as the eccentricity of a Newtonian orbit. It is seen that the case represented by $e^2 < 0$ would give a purely imaginary eccentricity in the Newtonian limit.

Two important distinctions that can be made among orbits are whether or not they terminate at $r = 0$ and whether or not they occur entirely within the Schwarzschild horizon. The $(e^2, s)$ parameter space can be divided into three regions which elucidate these differences. Non-terminating orbits are possible only within Region I (which also contains terminating orbits). Regions II and II' contain only terminating orbits with Region II' defining when the orbits occur within the horizon. Orbits can be unbounded when $e^2 \geq 1$.

As we have discussed briefly in ref.3, the boundaries of Regions I, II and II' can be obtained by extending the boundaries we presented in refs.1 and 2, and adding one for the left boundary of Region I. The three regions and their boundaries are shown in Fig.1. Region I is bounded by three curves: the $s = 0$ axis at the bottom which represents zero gravitational field, the $s = s_1(e)$ curve at the top that extends from a point which we call the vertex $V \equiv (e^2, s) = (-1/3, 1/2\sqrt{3})$ to $(\infty, 0)$, and the $s = s'_1(e)$ curve on the left that connects the vertex $V$ to the origin at $(0, 0)$, where $s_1(e)$ and $s'_1(e)$ are given by

$$s_1^2 = \frac{1 - 9e^2 + \sqrt{(1 - 9e^2)^2 + 27e^2(1 - e^2)^2}}{27(1 - e^2)^2}, \tag{6}$$

for $e \neq 1$, and $s_1^2 = 1/16$ for $e = 1$, and

$$s_1'^2 = \frac{1 - 9e^2 - \sqrt{(1 - 9e^2)^2 + 27e^2(1 - e^2)^2}}{27(1 - e^2)^2}. \tag{7}$$

It is seen that the extended parameter space of Region I characterized by $e^2 < 0$ is a small triangular-shaped space to the left of the $e^2 = 0$ axis between the two curves $s_1$ and $s_1'$ that intersect at the vertex $V$. The remaining part of the parameter space outside of Region I can be called Region II where there are only terminating orbits. It is convenient to separate out the region above the curve $s_2(e)$ given by

$$s_2^2 = \frac{1}{1 - e^2} \tag{8}$$

and call it Region II' where the particle's initial starting point is within the Schwarzschild horizon.

The left and upper boundary curves of Region I have a physical interpretation as the locations in the parameter space which represent all the circular orbits. When the periodic non-terminating solutions are evaluated along the left boundary curve $s_1'$ one obtains circular orbits having a range of radii

$$3\alpha \leq r \leq \infty \tag{9}$$

with $r = 3\alpha$ at the vertex point $V$ and $r = \infty$ at the origin. This range corresponds to the range of stable circular orbits with the innermost stable circular orbit (ISCO) occurring at the vertex point. All the finite circular orbits represented by $s_1'$ have $e^2 < 0$. No finite stable circular orbit exists with $e^2 \geq 0$. On $s_1'$ the value $e = 0$ occurs when $s = 0$ which implies zero gravitational field. Along the upper boundary $s_1$ the terminating orbits in Region I become circular with a range of radii

$$1.5\alpha \leq r \leq 3\alpha \tag{10}$$

where the lower limit is approached as $e^2$ goes to positive infinity and $r = 3\alpha$ at the vertex point $V$. This interval corresponds with the allowed sizes of the unstable circular orbits derived using the effective potential calculation. Evaluating the non-terminating solutions in Region I along $s_1$ gives the so-called asymptotic periodic orbits in which the limiting circular orbits have the same range of radii as the unstable circular orbits. The unstable circular orbits have $e^2 < 0$ over the interval $2.25\alpha < r \leq 3\alpha$ and $e^2 \geq 0$ over the range $1.5\alpha \leq r \leq 2.25\alpha$.

The boundaries of Region I can be expressed in terms of the modulus $k$ of the elliptic functions which are used to describe the exact solutions of the orbits. The upper boundary $s_1$ is defined by $k^2 = 1$ and the left boundary $s_1'$ has $k^2 = 0$. The value of $k^2$ is also zero along the $s = 0$ axis. One can also consider curves of constant $k^2$ for other allowed values between zero and one as they extend throughout the entire parameter space and we sketch some results in Fig. 2. Some physical quantities relating to the orbits such as the precession angle for periodic orbits do depend on $k$ so this description is of more than purely mathematical interest. It can be shown that all the lines of constant $k^2$ intersect the vertex point $V$. We also note the horizontal line representing $s = 1/2\sqrt{3}$ which intersects the vertex point. To the right of $V$ points on this line have a constant value of $k^2 = (1/2) + (\sqrt{3}/4) = 0.933012...$ To the left of $V$ points on the horizontal line take a constant value of $k^2 = (1/2) - (\sqrt{3}/4) = 0.0669872...$

In Region I the curves of constant $k^2$ for $e^2 \geq 0$ were described in refs.1 and 2. In the "extended" region where $e^2 < 0$ they turn upwards to meet the vertex point. Starting from the boundary curve $s_1'$ representing the stable circular orbits if one turns through Region I in a counterclockwise manner about the vertex point all values of $k^2$ from 0 to 1 are encountered in increasing order until reaching the boundary curve $s_1$ representing the unstable circular orbits.

The behavior of the curves of constant $k^2$ in Region II after they emerge from the vertex depends on the value of $k^2$. Starting from the boundary curve $s_1$ which represents the unstable circular orbits if one turns through Region II in a counterclockwise manner about the vertex point all values of $k^2$ from 1 to 0 are encountered in decreasing order until reaching the boundary curve $s_1'$ representing the stable circular orbits. Numerical calculations indicate that:

a) For $1 > k^2 > 0.933012...$ the curves of constant $k$ are below the horizontal $s = 1/2\sqrt{3}$ line to the right of $V$ and above the curve $s_1$ (where $k^2 = 1$) and decrease in value (for $s$) as $e^2 \to \infty$.

b) For $0.933012... > k^2 > 0.0669872...$ the curves are above the $s = 1/2\sqrt{3}$ line and eventually turn upwards. The line representing $k^2 = 1/2$ is a special case which can be expressed explicitly.

c) For $0.0669872... > k^2 > 0$ the curves are below the line defined by $s = 1/2\sqrt{3}$ to the left of $V$ and decrease in value (for $s$) as $e^2 \to -\infty$. For very small values of $k^2$ close to zero these curves fill the space below $s_1'$ before turning to the left.

Most of the analytical details related to the above description can be found in previous work but we mention some issues here.

The "combined" energy equation (3) can be rewritten as

$$\left(\frac{dU}{d\phi}\right)^2 = 4U^3 - g_2 U - g_3, \tag{11}$$

where

$$U = \frac{1}{4}\left(\frac{\alpha}{r} - \frac{1}{3}\right), \tag{12}$$

$$\begin{aligned} g_2 &= \frac{1}{12} - s^2, \\ g_3 &= \frac{1}{216} - \frac{1}{12}s^2 + \frac{1}{4}(1 - e^2)s^4. \end{aligned} \tag{13}$$

Region I is that sector of the parameter space $(e^2, s)$ where the discriminant

$$\Delta = 27g_3^2 - g_2^3 \tag{14}$$

of the cubic equation

$$4U^3 - g_2 U - g_3 = 0 \tag{15}$$

is less than or equal to zero. In this case the three roots of the cubic $e_1 \geq e_2 \geq e_3$ are all real. Regions II and II' have $\Delta > 0$ and in these regions the cubic equation has only one real root and two complex conjugate roots. When $\Delta = 0$ exactly at least two of the roots are equal which is a condition required for circular orbits.

The equation for the non-terminating orbits in Region I, bound and unbound, is given by [1]

$$\frac{1}{q} = \frac{1}{3} + 4e_3 + 4(e_2 - e_3)sn^2(\gamma\phi, k), \tag{16}$$

where $e_1 > e_2 \geq U > e_3$ and $q = r/\alpha$. The terminating orbit in Region I is given by [1]

$$\frac{1}{q} = \frac{1}{3} + 4\frac{e_1 - e_2 sn^2(\gamma\phi, k)}{cn^2(\gamma\phi, k)}. \tag{17}$$

where $U > e_1 > e_2 > e_3$. The constant $\gamma$ appearing in the argument and the modulus $k$ of the Jacobian elliptic functions [5] are given in terms of the three roots of the cubic equation (15) by

$$\begin{aligned} \gamma &= (e_1 - e_3)^{1/2}, \\ k^2 &= \frac{e_2 - e_3}{e_1 - e_3}, \end{aligned} \tag{18}$$

The roots $e_1, e_2, e_3$ are given by

$$\begin{aligned} e_1 &= 2\left(\frac{g_2}{12}\right)^{1/2} \cos\left(\frac{\theta}{3}\right), \\ e_2 &= 2\left(\frac{g_2}{12}\right)^{1/2} \cos\left(\frac{\theta}{3} + \frac{4\pi}{3}\right), \\ e_3 &= 2\left(\frac{g_2}{12}\right)^{1/2} \cos\left(\frac{\theta}{3} + \frac{2\pi}{3}\right), \end{aligned} \tag{19}$$

where

$$\cos\theta = g_3 \left(\frac{27}{g_2^3}\right)^{1/2}. \tag{20}$$

The roots can also be expressed as functions of $g_2$ and $k^2$ :

$$e_1 = \sqrt{\frac{g_2}{12}} \frac{2 - k^2}{\sqrt{1 - k^2 + k^4}}, \tag{21}$$

$$e_2 = \sqrt{\frac{g_2}{12}} \frac{-1 + 2k^2}{\sqrt{1 - k^2 + k^4}}, \tag{22}$$

$$e_3 = \sqrt{\frac{g_2}{12}} \frac{-1 - k^2}{\sqrt{1 - k^2 + k^4}}. \tag{23}$$

The modulus $k$ of the elliptic functions has a range $0 \leq k^2 \leq 1$ as $\theta$ varies from $0$ to $\pi$.

For a periodic type orbit the maximum distance $r_{\max}$ (for $-1/3 < e^2 < 1$) of a particle from the central object and its minimum distance $r_{\min}$, or their corresponding dimensionless forms $q_{\max}$ ($= r_{\max}/\alpha$) and $q_{\min}$ ($= r_{\min}/\alpha$), can be obtained from eq.(16) and they are given by

$$\frac{1}{q_{\max}} = \frac{1}{3} + 4e_3, \tag{24}$$

and

$$\frac{1}{q_{\min}} = \frac{1}{3} + 4e_2. \tag{25}$$

In contrast to the energy parameter $e$ defined by eq.(1), the true or geometrical eccentricity $\varepsilon$ of an bounded periodic orbit is defined by

$$\varepsilon = \frac{r_{\max} - r_{\min}}{r_{\max} + r_{\min}} = \frac{q_{\max} - q_{\min}}{q_{\max} + q_{\min}} = \frac{e_2 - e_3}{1/6 + e_2 + e_3}, \tag{26}$$

and the geometrical eccentricity $\varepsilon$ coincides with the energy parameter $e$ only when $s = 0$ for $0 \leq e < 1$ ($\varepsilon = 1$ when $e = 1$ for all values of $s$). The relationship between $e$ and $\varepsilon$ was exhibited in ref.1 for $0 \leq e \leq 1$.

It can be easily seen that in Region I when $e_2 = e_3$ the periodic solution (16) yields a circular orbit with $k^2 = 0$ and $\varepsilon = 0$. These circular orbits occur on $s = s_1'$ given by eq.(7) for $-1/3 \leq e^2 < 0$ (for which it is useful to note that $\theta = 0$ in eq.(20) and $\gamma = (1 - 12s_1'^2)^{1/4}/2$ from eq.(18)). When $e_1 = e_2$ the terminating solution (17) in Region I becomes a circle with $k^2 = 1$. The case $e_1 = e_2 = e_3 = 0$ represents the vertex point where the $k^2 = 0$ and $k^2 = 1$ curves meet. In Region I there is the relationship [1]

$$1 - 18s^2 + 54(1 - e^2)s^4 = \frac{(2 - k^2)(1 + k^2)(1 - 2k^2)}{2(1 - k^2 + k^4)^{3/2}}(1 - 12s^2) \tag{27}$$

which allows one to calculate the lines of constant $k^2$ in this region. When the coordinates of the vertex point $(e^2, s) = (-1/3, 1/2\sqrt{3})$ are substituted into the above equation one obtains a result which is true for all allowed values of $k^2$. All the lines of constant $k^2$ meet at the vertex point.

To express the terminating orbits in Regions II and II', we first define

$$A = \frac{1}{2}\left(g_3 + \sqrt{\frac{\Delta}{27}}\right)^{1/3},$$

$$B = \frac{1}{2}\left(g_3 - \sqrt{\frac{\Delta}{27}}\right)^{1/3}, \qquad (28)$$

where $g_3$ and $\Delta$ are defined by eqs.(13) and (14). The real root of the cubic equation (15) is given by

$$a = A + B \qquad (29)$$

and the two complex conjugate roots $b$ and $\bar{b}$ are $-(A+B)/2 \pm (A-B)\sqrt{3}i/2$. We further define

$$\gamma = [3(A^2 + AB + B^2)]^{1/4} \qquad (30)$$

and

$$k^2 = \frac{1}{2} - \frac{3(A+B)}{4\sqrt{3(A^2+AB+B^2)}} = \frac{1}{2} - \frac{3a}{4\gamma^2}. \qquad (31)$$

The terminating orbits in Regions II and II' are then given by [1]

$$\frac{1}{q} = \frac{1}{3} + 4a + 4\gamma^2 \frac{1 - cn(2\gamma\phi, k)}{1 + cn(2\gamma\phi, k)}. \qquad (32)$$

At the vertex point $(e^2, s) = (-1/3, 1/2\sqrt{3})$ one can calculate that $g_2 = g_3 = 0$ so that $A = B = 0$. Therefore $a = \gamma = 0$ and the relationship

$$4\gamma^2 k^2 = 2\gamma^2 - 3a$$

derived from eq.(31) holds for all values of $k^2$, i.e. the lines of constant $k^2$ in Region II meet at the vertex point. Along the horizontal line $s = 1/2\sqrt{3}$ one has $g_2 = 0$ so that $\Delta = 27g_3^2$. In this case $A = (2g_3)^{1/3}/2$, $B = 0$, $a = A$, and $\gamma = (3A^2)^{1/4}$. It can be checked that $g_3 < 0$ when $e^2 > -1/3$ and $s^2 = 1/12$ so that $A < 0$ and

$$k^2 = \frac{1}{2} - \frac{\sqrt{3}A}{4|A|} = \frac{1}{2} + \frac{\sqrt{3}}{4} = 0.933012... \qquad (33)$$

Similarly when $e^2 < -1/3$ and $s^2 = 1/12$, one has $g_3 > 0$ so $A > 0$ and

$$k^2 = \frac{1}{2} - \frac{\sqrt{3}}{4} = 0.0669872... \qquad (34)$$

This establishes the results previously described for the horizontal line defined by $s = 1/2\sqrt{3}$ which represents $g_2 = 0$.

The curve represented by $k^2 = 1/2$ is given by

$$s^2 = \frac{1}{6(1-e^2)}\left(1 - \sqrt{\frac{1+2e^2}{3}}\right), \tag{35}$$

for $e^2 \geq -1/2$, $0 \leq s^2 \leq 1/9$ (in Region I and part of Region II), and it is given by

$$s^2 = \frac{1}{6(1-e^2)}\left(1 + \sqrt{\frac{1+2e^2}{3}}\right), \tag{36}$$

for $e^2 \geq -1/2$, $s^2 \geq 1/9$ (in Region II). Together, the entire $k^2 = 1/2$ curve represents $g_3 = 0$. The intersection of $g_2 = 0$ and $g_3 = 0$ curves is the vertex point $V$.

On the vertical line $e^2 = -1/3$ through the vertex point $V$, the squared modulus $k^2$ as a function of $s$ can be shown to be given by

$$k^2 = \frac{1}{2} \mp \frac{\sqrt{3}}{4} \frac{1 + (1-12s^2)^{1/3}}{\sqrt{1 + (1-12s^2)^{1/3} + (1-12s^2)^{2/3}}}, s^2 \lessgtr \frac{1}{12}. \tag{37}$$

The periods of the elliptic functions appearing in the orbital equations (16), (17), and (32) depend on the complete elliptic integral of the first kind $K(k)$ so orbital parameters relating to angles depend on the modulus $k$. For example, the precession angle of the bounded periodic orbits is

$$\Delta\phi = \frac{2K(k)}{\gamma} - 2\pi \tag{38}$$

where $\gamma$ is given by (18). For terminating orbits that begin at a finite distance from the central object and have initial velocity perpendicular to a line joining the central object to the orbiting particle the total turning angle that the particle makes when $r = 0$ is $K(k)/\gamma$ where $\gamma$ is given by (18) for Region I and by (30) for Region II. For the bounded periodic orbits the roots $e_1, e_2, e_3$ can be simply described in terms of $k^2$ and $s^2$ (see eqs. (21), (22), and (23)) which demonstrates that the maximum and minimum distances of the orbiting particle from the central object also have a functional dependence on $k$. In particular, for circular orbits characterized by $\varepsilon = 0$, $k^2 = 0$, $\theta = 0$ that occur on $s = s_1'$ given by eq.(7), the precession angle is given by

$$\Delta\phi = 2\pi[(1 - 12s_1'^2)^{-1/4} - 1] \tag{39}$$

for $0 \leq s_1'^2 \leq 1/12$. In terms of the radius $r$ of a stable circular orbit in the range given by eq.(9), the precession angle is given by

$$\Delta\phi = 2\pi\left\{\left[1 - \frac{6\alpha}{r^2}\left(r - \frac{3\alpha}{2}\right)\right]^{-1/4} - 1\right\}.$$

Thus excluding the case of zero gravitational field $s_1' = 0$ for which the circular orbits have an infinite radius, all (perfectly) circular orbits have a non-zero precession angle. The precession angle for the so-called ISCO, i.e. the circular orbit that occurs at (the vertex point $V$) $s_1'^2 = 1/12$, $e^2 = -1/3$, is infinite.

For the parabolic-type orbits characterized by $e = 1$ which coincides with $\varepsilon = 1$ (see ref.[1] for proof) in Region I ($0 \leq s^2 \leq 1/16$), $k$ and $\gamma$ can be expressed explicitly as a function of $s$ by

$$k^2 = \frac{1 - 8s^2 - \sqrt{1 - 16s^2}}{8s^2}, \gamma^2 = \frac{2s^2}{1 - \sqrt{1 - 16s^2}}, \tag{40}$$

and they can be used to find the precession angle $\Delta\phi$ from eq.(38) that represents the angle between the incoming and outgoing trajectories of the orbit.

In summary the investigation of the parameter space $(e^2, s)$ has revealed at least three interesting features:

1) All circular orbits occur on the upper and left boundaries of Region I which contains the non-terminating trajectories. These boundaries can be described mathematically either by the explicit equations for $s_1$ and $s_1'$ given in eqs. (6) and (7), or as the curves of constant $k^2 = 1$ and $k^2 = 0$, or by the condition $\Delta = 0$ where $\Delta$ is the discriminant of the cubic expression (see eqs. (11) and (15)) occurring in the energy equation ($\Delta = 0$ and $k^2 = 0$ are also satisfied by the trivial $s = 0$ axis).

All stable circular orbits with $\varepsilon = 0$ are on the left boundary of Region I and are in the region $e^2 < 0$ that does not have a Newtonian correspondence. For example, all perfectly circular orbits have a non-zero precession angle and thus require a general relativistic consideration.

2) All the curves of constant $k^2$ intersect the vertex point $V$ representing the innermost stable circular orbit for which $e^2 = -1/3$ and $s = 1/2\sqrt{3}$. In the vicinity of $V$ these curves appear to radiate outwards from $V$ into Region II in decreasing order of $k^2$ in a counterclockwise sense.

3) In Region II the horizontal line defined by $s = 1/2\sqrt{3}$ also has constant values of $k^2 = 0.933012...$ to the right of $V$ and $k^2 = 0.0669872...$ to the left of $V$. This line appears to delineate regions of parameter space in which the curves of constant $k^2$ have different behaviors.

Finally, since $g_2$ and $g_3$ depend on $s^2$, not on $s$, most of the expressions of interest derived in our analysis of the orbits depend only on $s^2$. For this reason it may be advantageous to use the parameters $(e^2, s^2)$ in some cases when studying these trajectories.

Acknowledgement
We are grateful to Christopher Berry for many helpful comments.

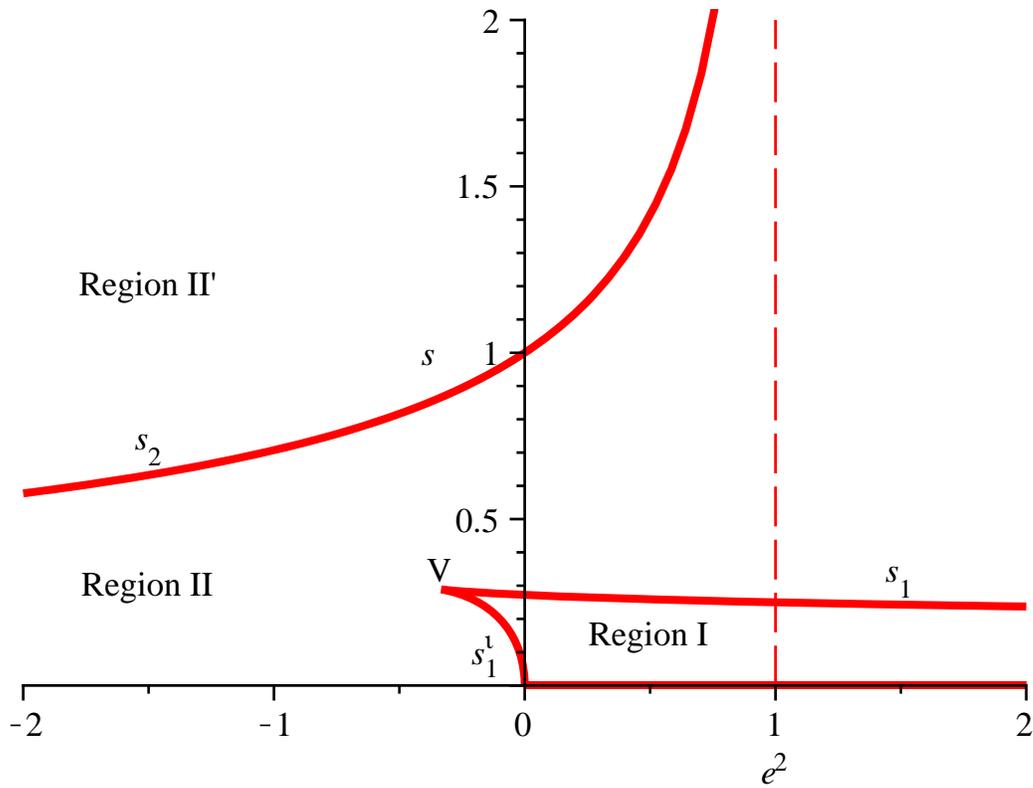

Fig.1 The boundaries of Regions I, II, and II' are displayed. The vertex point V occurs at the intersection of the left boundary $s_1^{\iota}$ and the upper boundary $s_1$. Circular orbits are represented on these two boundaries of Region I with the innermost stable circular orbit occurring at V. Unbound orbits are possible when $1 \leq e^2$.

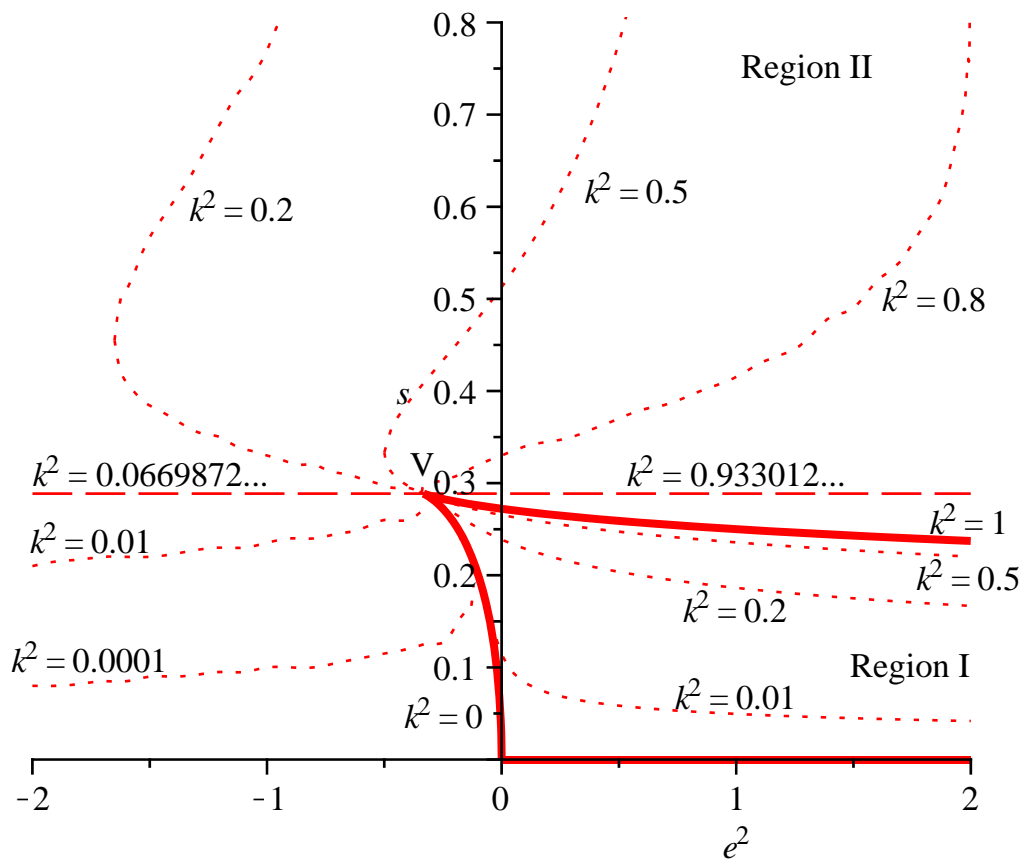

Fig.2 Curves of constant $k^2$. Within Region I only three values of $k^2$ are displayed. All curves intersect the point V. The dashed horizontal line through V has $s^2 = \dfrac{1}{12}$ .